\documentstyle[preprint,aps]{revtex}
\begin{document}
\newcommand{\cross}{\mbox{$\rlap{\kern0.125em/}$}}
\draft
\preprint{\vbox{\hbox{IFT--P.026/98}}}
\title{Constraints on Electroweak Contact Interactions from LEP and 
Tevatron Data.}
\author{Alexander Belyaev $^{1,2}$ and Rogerio Rosenfeld$^1$}
\address{ 
$^1$ {\it Instituto de F\'\i sica Te\' orica, Universidade 
Estadual Paulista},\\
Rua Pamplona 145, 01405-900 - S\~ao Paulo, S.P., Brasil\\
$^2$ {\it Skobeltsin Institute for Nuclear Physics,
Moscow State University, \\ 119 899, Moscow, Russian Federation}
}
\maketitle
\widetext
\vspace*{0cm}
\begin{abstract}
A complete set of dimension-6 effective contact interactions involving Higgs, 
gauge bosons and quarks is studied. Limits on the coefficients of these
new operators are obtained from the experimental values of the 
$Z$ and $W$ gauge bosons widths.
\end{abstract}
\vspace*{0cm}
\vspace*{0.5cm}
\section{Introduction} 
\label{int}
There have been extensive theoretical and phenomenological studies of physics beyond 
the Standard Model (SM) in recent years. 
This new physics could be revealed, for example, in studies
of anomalous couplings that are either different or absent at tree level
in the SM.
Studies for such couplings include searches for anomalous gauge boson 
self-couplings,
Higgs boson couplings and top quark couplings.

The most general phenomenological parametrization for new phenomena
beyond the SM  can be achieved by means of an effective
Lagrangian \cite{effective} that involves  operators with
dimension higher than four, containing the relevant fields at
low energies and respecting the symmetries of the Standard Model.  
The effective Lagrangian approach is a model--independent
way to describe  new physics that can occur at an energy scale
$\Lambda$ much larger than the scale where the experiments  are
performed. New particles present at the energy scale $\Lambda$
are integrated out at low energies, generating higher dimensional
operators.
The effective Lagrangian depends on the particle content at low
energies and here we assume that the Higgs boson can be light,
being present in the higher dimensional operators, in addition to
the electroweak gauge bosons and fermions, and the SM symmetries are linearly
realized \cite{linear,hisz}. In what follows we adopt the notation
of  Buchm\"uller and Wyler in ref. \cite{linear}.

Limits on operators modifying the 
electroweak and Higgs bosons interactions but leaving the fermion interactions
untouched have been obtained
from studying many different processes 
\cite{hagiwara2,e,our,gamma,stirling,contact,serg1,serg2}.
In this short note we concentrate on possible contact interactions involving 
Higgs, gauge bosons {\it and} quarks.  These interactions are
interesting since they could in principle contribute to 
associated Higgs and gauge boson production at hadron machines and
to anomalous top quark couplings. In fact,  
some of the analogous leptonic 
contact operators have been 
used in the study of anomalous Higgs production at LEP2, like
${\cal O}_{\Phi l}^{(1)}, {\cal O}_{\Phi l}^{(3)}$ and 
${\cal O}_{\Phi e}$ \cite{contact}. Also the operators like 
${\cal O}_{\Phi q}^{(1)}$,${\cal O}_{\Phi q}^{(3)}$,${\cal O}_{D u}$,
${\cal O}_{D d}$,${\cal O}_{u W}$ and ${\cal O}_{d W}$  have been studied in 
the context of anomalous top
quark production at the Tevatron \cite{top} and $e-\gamma$ colliders \cite{eg}.

These operators are constrained at tree-level
by the decay widths of the $Z$ and $W$ boson.
In this paper we obtain limits for each
coefficient of a complete set of contact interactions involving 
Higgs, gauge bosons {\it and} quarks, assuming that different generations 
can have different anomalous couplings.  We also write down the specific linear
combination of coefficients that are bounded in each process and notice that
accidental cancellations could avoid these direct bounds.

\section{Effective Lagrangian and Contact Interactions }
\label{eff:lag}

The most general effective
Lagrangian containing dimension--6 terms describing
contact interactions among electroweak gauge fields, {\it i.e.\/}
$\gamma$, $W^{\pm}$, $Z^0$, Higgs boson $H$, and quark fields is
given by \cite{linear}:
\begin{eqnarray}
{\cal L}_{\text{eff}} &=& {\cal L}_{\text{SM}} + \frac{1}{\Lambda^2} \left(  
f_{\Phi q}^{(1)} {\cal O}_{\Phi q}^{(1)} + f_{\Phi q}^{(3)} {\cal O}_{\Phi q}^{(3)}
+ f_{\Phi u} {\cal O}_{\Phi u} + \right. \\ \nonumber
& & f_{\Phi d} {\cal O}_{\Phi d} +
f_{\Phi \Phi} {\cal O}_{\Phi \Phi} + 
f_{D u} {\cal O}_{D u} + f_{\bar{D} u} {\cal O}_{\bar{D} u} +
f_{D d} {\cal O}_{D d} + f_{\bar{D} d} {\cal O}_{\bar{D} d} + \\ \nonumber
& & \left. f_{u W} {\cal O}_{u W} +f_{u B} {\cal O}_{u B} +
f_{d W} {\cal O}_{d W} +f_{d B} {\cal O}_{d B}  \right) + \mbox{h.c.}
\label{lagrangian}
\end{eqnarray}
with each operator ${\cal O}_i$ defined as, 
\begin{eqnarray}
{\cal O}_{\Phi q}^{(1)} &=& i ( \Phi^{\dagger} D_{\mu} \Phi ) 
                            (\bar{q} \gamma^{\mu} q) \\
{\cal O}_{\Phi q}^{(3)} &=& i ( \Phi^{\dagger} D_{\mu} \tau^I \Phi ) 
                            (\bar{q} \gamma^{\mu} \tau^I q) \\
{\cal O}_{\Phi u} &=& i ( \Phi^{\dagger} D_{\mu} \Phi ) 
                            (\bar{u} \gamma^{\mu} u) \\
{\cal O}_{\Phi d} &=& i ( \Phi^{\dagger} D_{\mu} \Phi ) 
                            (\bar{d} \gamma^{\mu} d) \\
{\cal O}_{\Phi \Phi} &=& i ( \Phi^{\dagger} \varepsilon D_{\mu} \Phi ) 
                            (\bar{u} \gamma^{\mu} d) \\
{\cal O}_{D u} &=& (\bar{q} D_{\mu} u) (D^{\mu} \tilde{\Phi} ) \\                            
{\cal O}_{\bar{D} u} &=& ( D_{\mu}\bar{q} u) (D^{\mu} \tilde{\Phi} ) \\ 
{\cal O}_{D d} &=& (\bar{q} D_{\mu} d) (D^{\mu} \Phi ) \\  
{\cal O}_{\bar{D} d} &=& ( D_{\mu}\bar{q} d) (D^{\mu} \Phi ) \\
{\cal O}_{u W} &=& (\bar{q} \sigma^{\mu \nu} \tau^I u) \Phi W_{\mu \nu}^I \\
{\cal O}_{d W} &=& (\bar{q} \sigma^{\mu \nu} \tau^I d) \Phi W_{\mu \nu}^I \\
{\cal O}_{u B} &=& (\bar{q} \sigma^{\mu \nu}  u) \Phi B_{\mu \nu} \\
{\cal O}_{d B} &=& (\bar{q} \sigma^{\mu \nu}  d) \Phi B_{\mu \nu} \; ,
%
%
\end{eqnarray}
where $\Phi$ is the Higgs field doublet, which in the unitary
gauge assumes the form,
\[
\Phi = \left(\begin{array}{c}
0 \\
(v + H)/\sqrt{2}
\end{array}
\right) \; , 
\]
with $B_{\mu \nu}$ and $ W^I_{\mu \nu}$ being the field strength
tensors of the $U(1)$ and $SU(2)$ gauge fields respectively.

We incorporated all the new interactions arising from the effective 
lagrangian 
in the Comphep package \cite{comphep} by using the code Lanhep \cite{lanhep}.

For the $Z$ boson, we obtain the  modified  $Zb\bar b$  and $Zc\bar c$
vertices which can be written as follows:

\begin{eqnarray}
 \Gamma_{\mu}^{Zb\bar{b}} &=&
 i \frac{e}{2 s_Wc_W}  
 \left[         
 \gamma^\mu (g_V^d + g_A^d \gamma_5)+
 \gamma^\mu (\kappa_1^{dV}-\kappa_1^{dA}\gamma_5)/2 + 
 (p_1^\mu- p_2^\mu) \kappa_2^{dV}/M_Z+\right.\nonumber\\
&&\left. (p_1^\mu+ p_2^\mu)\gamma_5\kappa_2^{dA}/M_Z+
(i \sigma^{\mu \nu}p_{Z \nu}) \kappa_3^d/M_Z \right]
\end{eqnarray}
 where
\begin{eqnarray} 
g_V^d     &=&  -1/2 + 2/3 s_w^2   \nonumber\\                                                     
g_A^d     &=&  -1/2                                                   
\end{eqnarray}
and
\begin{eqnarray} 
\kappa_1^{dV}&=& -4\frac{M_Z^2}{e^2\Lambda^2} (f_{\Phi q}^{(1)} + f_{\Phi q}^{(3)}
 + f_{\Phi d})s_{w}^2  c_{w}^2  \\
\kappa_1^{dA}&=& -4\frac{M_Z^2}{e^2\Lambda^2} (f_{\Phi q}^{(1)} + f_{\Phi q}^{(3)}
 - f_{\Phi d})s_{w}^2  c_{w}^2   \\
 \kappa_2^{dV}&=& 
  \frac{M_Z^2}{e \Lambda^2} (-f_{D d} + f_{\bar{D} d})s_{w}  c_{w} /\sqrt{2} \\
\kappa_2^{dA}&=& 
  \frac{M_Z^2}{e \Lambda^2} (f_{D d} + f_{\bar{D} d})s_{w}  c_{w} /\sqrt{2}\\
\kappa_3^d&=&4 \sqrt{2}\frac{M_Z^2}{ e^2 \Lambda^2}  ( f_{d W} c_{w}+f_{d B} s_{w})
s_{w}^2  c_{w}^2
\end{eqnarray}
and
\begin{eqnarray} 
\Gamma_{\mu}^{Zc\bar{c}} &=&                                                                                                                                                                                                                                   
i \frac{e}{2 s_W c_W}                              
 \left[         
\gamma^\mu (g_V^u + g_A^u \gamma_5)-
\gamma^\mu(\kappa_1^{uV}-\kappa_1^{uA} \gamma_5)/2 - 
(p_1^\mu-p_2^\mu)\kappa_2^{uV}/M_Z-\right.\nonumber\\
&&\left.(p_1^\mu+p_2^\mu)\gamma_5 \kappa_2^{uA}/M_Z- 
(i \sigma^{\mu \nu}p_{Z \nu}) \kappa_3^{u}/M_Z \right]																																																																						
\end{eqnarray}
where
\begin{eqnarray} 
g_V^u     &=&  1/2 - 4/3 s_w^2   \nonumber\\                                   
g_A^u     &=&  1/2   
\end{eqnarray}
and 
\begin{eqnarray}
\kappa_1^{uV}&=& 3\frac{M_Z^2}{e^2\Lambda^2} (f_{\Phi q}^{(1)} - f_{\Phi q}^{(3)}
 + f_{\Phi u}) s_{w}^2  c_{w}^2 \\
\kappa_1^{uA}&=& 3\frac{M_Z^2}{e^2\Lambda^2} (f_{\Phi q}^{(1)} - f_{\Phi q}^{(3)}
 - f_{\Phi u})s_{w}^2  c_{w}^2  \\
 \kappa_2^{uV}&=& 
 \frac{M_Z^2}{e\Lambda^2} (-f_{D u} + f_{\bar{D} u}) \cos (2 \theta_W)s_{w} c_{w}/\sqrt{2}\\
\kappa_2^{uA}&=& 
  \frac{M_Z^2}{e \Lambda^2} (f_{D u} + f_{\bar{D} u}) \cos (2 \theta_W)s_{w} c_{w}/\sqrt{2}\\
\kappa_3^u&=& 4 \sqrt{2}\frac{M_Z^2}{e^2 \Lambda^2}  ( f_{u W} c_{w}-f_{u B} s_{w})
s_{w}^2  c_{w}^2
\end{eqnarray}
 
The vertex $W^- u\bar d$  can be written as follows:
\begin{eqnarray} 
\Gamma_{\mu}^{W^- u\bar{d}} &=& 
i \frac{e V_{ud}}{2\sqrt{2}s_W} 
 \left[         
(1-\gamma_5) \gamma_\mu + 
\gamma_\mu (\kappa_4^{udV}-\kappa_4^{udA} \gamma_5)                                   
-( \kappa_1^{udV} - \kappa_1^{udA}\gamma_5)  p_{\bar d}^\mu /M_W
\right.\nonumber\\
&&\left.
-(\kappa_2^{udV}  - \kappa_2^{udA}\gamma_5) p_{u}^\mu/M_W
-(\kappa_3^{udV} + \kappa_3^{udA}\gamma_5) i \sigma^{\mu \nu} p_{W \nu}/ M_W
\right]	                                                                                                                                                                                                                                                       
 \end{eqnarray}                                                                                                                                                                                                                                                               
where
\begin{eqnarray}
\kappa_1^{udV}&=&\frac{ \sqrt{2}s_{w} M_W^2}{e \Lambda^2} (f_{\bar{d}u} - f_{Dd} ) \\
\kappa_1^{udA}&=&\frac{\sqrt{2} s_{w} M_W^2}{e \Lambda^2} (f_{\bar{d}u} + f_{Dd} ) \\
\kappa_2^{udV}&=&\frac{\sqrt{2} s_{w} M_W^2}{e \Lambda^2} (f_{Du} - f_{\bar{d}d} ) \\
\kappa_2^{udA}&=&\frac{\sqrt{2} s_{w} M_W^2}{e \Lambda^2} (f_{Du} + f_{\bar{d}d} ) \\
\kappa_3^{udV}&=&\frac{4 \sqrt{2}s_{w} M_W^2}{e^2\Lambda^2}(f_{u W}  +f_{d W} ) s_w \\
\kappa_3^{udA}&=&\frac{4 \sqrt{2}s_{w} M_W^2}{e^2\Lambda^2}(f_{u W}  -f_{d W} ) s_w \\
\kappa_4^{udV}&=&\frac{2 s_{w}^2 M_W^2}{e^2\Lambda^2}(f_{\Phi \Phi}  + 2 f_{\Phi q}^3 ) \\
\kappa_4^{udA}&=&\frac{2 s_{w}^2 M_W^2}{e^2\Lambda^2}(f_{\Phi \Phi}  - 2 f_{\Phi q}^3 ) 
\end{eqnarray}

The results obtained for the $Z$ decay width into $u$- and $d$-type and for 
$W$ decay into quarks are shown in the Appendix, where we include 
finite quark mass effects
and the appropriate QCD and QED radiative corrections.

From these expressions we note some interesting features: \\
$\bullet$ Only $ \kappa_1^{u(V,A)}$, $\kappa_1^{d(V,A)}$ and
$\kappa_3^{ud(V,A)}$ interfere with 
the Standard Model
result. The other combinations only appear quadratically and therefore should
be less constrained by data. \\
$\bullet$ There could be cancellations in the linear combinations of the 
coefficients of the effective lagrangian that would the widths less
sensitive to them, but we don't consider this possibility further. \\
$\bullet$ The operator ${\cal O}_{\Phi \Phi}$ is not constrained by $Z$ decays.

We get bounds on the different coefficients by requiring  
$3 \sigma$ deviation criteria for  $Z$ and $W$ decay widths.
For the $Z$ boson, we consider deviations in the total hadronic width
due to anomalous couplings of the first family only and also
deviations in $R_c$ and $R_b$ arising from anomalous couplings of the
second and third family respectively. For the $W$ boson we considered
deviations in the total width due to anomalous couplings in the
first generation. We used the 
data \cite{pdg}:
\begin{equation}
\Gamma(Z)_{\mbox{hadrons}} = 1.7432 \pm 0.0023 \;\mbox{GeV} ; \;\;\;
R_b = 0.2170 \pm 0.0009 ; \;\;\;
R_c = 0.1734 \pm 0.0048
\end{equation}
and
\begin{equation}
\Gamma(W)_{\mbox{total}} = 2.07 \pm 0.06 \;\mbox{GeV} ; \;\;\;
\end{equation}

The limits obtained are presented in Tables I and II.
We clearly see a hierarchy in the constraints: 
the coefficients $f_{\Phi q}^{(1,3)}$,
$f_{\Phi (u,d)}$ are tightly bounded to be less
than $1 \; \mbox{TeV}^{-2}$ in most cases, $f_{(u,d) (W,B)}$ are 
roughly less than a few $ \; \mbox{TeV}^{-2}$, whereas 
$f_{\Phi \Phi}$, $f_{D (u,d)}$, $f_{\bar{D} (u,d)}$ are typically
constrained to be less than $ 20 \; \mbox{TeV}^{-2}$.

We would like to stress that the
limits shown  below (especially for the third quark's generation)
exclude several operators 
for anomalous top-quark and Higgs physics with  precision which
is out of range for the present and future hadron colliders.
For example, limits on the
couplings  $f_{\Phi q}^{(1,3)}$, $f_{D d}$,  $f_{\bar{D} d}$ and 
$f_{dW}$ 
bound different structures of $Wtb$ anomalous coupling, as also noted 
recently in \cite{eg} (however, apparently they did not consider all
possible structures in the interaction vertices).
All couplings we considered are also related to 
anomalous Higgs interactions with quarks and
gauge bosons like $Hqq$, $HVqq$. Therefore the operators considered 
in this paper should be taken into account if one study anomalous
Higgs couplings effects at the colliders.

\section{Conclusions}
\label{con}

The search for the effect of higher dimensional operators that
give rise to anomalous couplings should be pursued
in all possible processes since the results may provide important
information on physics beyond the Standard Model.
However, one should not forget that the contribution of some of these 
operators are strongly bounded due to precision measurements at LEP.

In this note we derived limits on the coefficients of a complete set of
dimension-6 operators involving electroweak gauge bosons, quarks and Higgs 
fields. We showed the linear combinations of these operators that are
important for the decay processes. In particular, there are some operators
that are not tightly bound by the gauge boson decay widths but they could be
important for other processes like anomalous associated Higgs production at 
the Tevatron.

\acknowledgments
One of the authors (A.S.B.) is grateful to A. V. Semenov for important
improvements of the LanHEP package. This work was supported by Conselho Nacional de Desenvolvimento
Cient\'{\i}fico e Tecnol\'ogico (CNPq), and by Funda\c{c}\~ao de
Amparo \`a Pesquisa do Estado de S\~ao Paulo (FAPESP).

\newpage
\begin{table}
\begin{tabular}{|| l || l | l | l ||}
 Anomalous  Couplings&\multicolumn{3}{c ||}{ Limits}   \\
\hline
&
\multicolumn{1}{ c |}{1-st generation }&
\multicolumn{1}{ c |}{2-nd generation }&
\multicolumn{1}{ c ||}{3-rd generation }\\
\hline 
\hline
$f_{\Phi q}^{(1)}/\Lambda^2$   
&( $-$0.17, 0.17 )   
&( $-$0.56, 0.58 )	
&( $-$0.092, 0.092 )\\
\hline
$f_{\Phi q}^{(3)}/\Lambda^2$   
& ( $-$0.10, 0.10 )
& ( $-$0.59, 0.56 )
& ( $-$0.093, 0.092 )     \\
\hline
$f_{\Phi u}/\Lambda^2$   
& ( $-$0.54, 0.54 )
& ( $-$1.6, 1.2 )
& -----      \\
\hline
$f_{\Phi d}/\Lambda^2$   
& ( $-$0.65, 0.90 )
& -----
& ( $-$0.50, 0.63 ) \\
\hline
$f_{\Phi \Phi}/\Lambda^2$   
 & -----	 & -----	 & -----	 \\
\hline
$f_{D u}/\Lambda^2$   
& ( $-$26, 26 )
& ( $-$52, 49 )
& -----	       \\
 \hline
$f_{D d}/\Lambda^2$   
& ( $-$14, 14 )
& -----
& ( $-$17, 8.8 ) \\
\hline
$f_{\bar{D} u}/\Lambda^2$   
& ( $-$26, 26 )
& ( $-$49, 52 )
& -----	       \\
\hline
$f_{\bar{D} d}/\Lambda^2$   
& ( $-$14, 14 )
& -----
& ( $-$8.8, 17 ) \\
\hline
$f_{u W}/\Lambda^2$   
& ( $-$1.6, 1.6 )
& ( $-$3.2, 2.7 )
& ----- \\
\hline
$f_{u B}/\Lambda^2$   
& ( $-$2.9, 2.9 )
& ( $-$5.1, 6.0 )
& ----- \\
\hline
$f_{d W}/\Lambda^2$   
& ( $-$1.6, 1.6 )
& -----
& ( $-$3.2, 0.53 ) \\
\hline
$f_{d B}/\Lambda^2$   
& ( $-$2.9, 2.9 )
& -----
& ( $-$1.9, 0.98 ) 
\end{tabular}
\caption{Intervals of allowed values of the
coefficients $f_i/\Lambda^2$ in units of TeV$^{-2}$  using  
$3 \,\sigma$ deviation criteria
for $\Gamma(Z \rightarrow hadrons)$ ($\Gamma(Z \rightarrow \bar{c} c)$)
for coefficients of the 1-st and second quark's generation and 
for $\Gamma(Z \rightarrow \bar{b} b)$
for coefficients of the 3-rd quark's generation.}
\label{tab:1}
\end{table}

\begin{table}
\begin{tabular}{|| l || l ||}
 Anomalous  Couplings& Limits   \\
\hline 
\hline
$f_{\Phi \Phi}/\Lambda^2$   
& ( $-$17, 17) \\
\hline
$f_{D u}/\Lambda^2$   
& ( $-$104, 104 ) \\
 \hline
$f_{D d}/\Lambda^2$   
& ( $-$104, 104 ) \\
\hline
$f_{\bar{D} u}/\Lambda^2$   
& ( $-$104, 104 ) \\
\hline
$f_{\bar{D} d}/\Lambda^2$   
& ( $-$104, 104) \\
\hline
$f_{u W}/\Lambda^2$   
& ( $-$8.7, 8.7 ) \\
\hline
$f_{d W}/\Lambda^2$   
& ( $-$8.7, 8.7 ) 
 \end{tabular}
\caption{Intervals of allowed values of the
coefficients $f_i/\Lambda^2$ in units of TeV$^{-2}$  using  
$3 \,\sigma$ deviation criteria
for $\Gamma(W)_{\mbox{total}}$ .}
\label{tab:2}
\end{table}

\newpage
\section*{Appendix}
\setcounter{equation}{0}
In this appendix we collect the results for the widths, including finite
mass effects and radiative corrections. 

The $Z$ partial decay width into
$u-$type quarks is given by:

\begin{eqnarray}
\Gamma (Z\rightarrow U\bar{U})&=&
\frac{e^2 M_Z}{64\pi c_{w}^2  s_{w}^2}\sqrt{(1-4 M_U^2/M_Z^2)}
\left[
4 ((g_A^u)^2 + (g_V^u)^2) - 4 (g_V^u  \kappa_1^{uV} - g_A^u  \kappa_1^{uA}) +\right. \nonumber\\
&& \left. 
(\kappa_1^{uA})^2 + (\kappa_1^{uV})^2 + 2 (\kappa_2^{uV})^2 - 4 \kappa_2^{uV} \kappa_3^{u} + 2 (\kappa_3^{u})^2 + \Delta_U \right]
\end{eqnarray}

Finite quark mass effects are described by the
function:
\begin{eqnarray}
\Delta_U &=&  \left[
32 (\kappa_2^{uV})^2 M_U^4/M_Z^4 + 
16 (2 g_V^u  - \kappa_1^{uV}) \kappa_2^{uV} M_U^3/M_Z^3 +\right. \nonumber \\
& & \left.
2 ( 4(g_V^u)^2 - 8 (g_A^u)^2 + 4 (2 g_A^u  \kappa_1^{uA} - g_V^u  \kappa_1^{uV})- 
2 (\kappa_1^{uA})^2  + (\kappa_1^{uV})^2 - \right. \nonumber \\
& & \left.
8 ((\kappa_2^{uV})^2 - \kappa_2^{uV} \kappa_3^{u} - (\kappa_3^{u})^2) ) M_U^2/M_Z^2 +
4 (-2 g_V^u  + \kappa_1^{uV}) (\kappa_2^{uV} - 3 \kappa_3^{u}) M_U/M_Z \right]
\end{eqnarray}

The $Z$ partial decay width into
$d-$type quarks is given by:

\begin{eqnarray}
\Gamma (Z\rightarrow D\bar{D})&=&
\frac{e^2 M_Z}{64\pi c_{w}^2  s_{w}^2}\sqrt{(1-4 M_D^2/M_Z^2)}
\left[
4 ((g_A^d)^2 + (g_V^d)^2) + 4 (g_V^d  \kappa_1{dV} + g_A^d  
\kappa_1^{dA}) +\right. \nonumber\\
&& \left. 
(\kappa_1^{dA})^2 + (\kappa_1^{dV})^2 + 2 (\kappa_2^{dV})^2 - 
4 \kappa_2^{dV} \kappa_3^{d} + 2 (\kappa_3^{d})^2 + \Delta_D \right]
\end{eqnarray}

Finite quark mass effects are described by the
function:
\begin{eqnarray}
\Delta_D &=&  \left[
32 (\kappa_2^{dV})^2 M_D^4/M_Z^4 - 
16 (2 g_V^d  + \kappa_1^{dV}) \kappa_2^{dV} M_D^3/M_Z^3 +\right. \nonumber \\
& & \left.
2 ( 4(g_V^d)^2 - 8(g_A^d)^2 + 4(-2 g_A^d  \kappa_1^{dA} + g_V^d  \kappa_1^{dV})- 
2 (\kappa_1^{dA})^2  + (\kappa_1^{dV})^2 - \right. \nonumber \\
& & \left.
8 ((\kappa_2^{dV})^2 - \kappa_2^{dV} \kappa_3^{d} - (\kappa_3^{d})^2) ) M_D^2/M_Z^2 +
4 (2 g_V^d  + \kappa_1^{dV}) (\kappa_2^{dV} - 3 \kappa_3^{d}) M_D/M_Z \right]
\end{eqnarray}

The $W$-boson partial decay width into
$\bar d$ and $u$ type quarks is given by: 
and for W decay:
\begin{eqnarray}
\Gamma (W^+\rightarrow u\bar{d})&=&
\frac{e^2 V_{ud}^2}{256\pi} \frac{M_{W}}{s_{w}^2}
\left[8(1 + \kappa_4^{udV})^2 +8(1 + \kappa_4^{udA})^2 
 + \right. \nonumber\\
&& \left. 
(-\kappa_1^{udV}+\kappa_2^{udV}+ 2 \kappa_3^{udV})^2 + 
(-\kappa_1^{udA}+\kappa_2^{udA}+ 2 \kappa_3^{udA})^2
\right] .
\end{eqnarray}

We incorporate the appropriate QCD and QED radiative corrections by including 
the factors:

\begin{eqnarray}
\delta_{QCD} &=& 1 + \frac{\alpha_s}{\pi} + 1.409 \frac{\alpha_s^2}{\pi^2} -
12.77 \frac{\alpha_s^3}{\pi^3} = 1.03954   \\
\delta_{QED} &=& 1 + \frac{3 \alpha Q_f^2}{4 \pi} \\
\delta_{b\bar{b}} &=& 1 + 0.01 (- \frac{m_t^2}{2 M_Z^2} + \frac{1}{5} )
= 0.98359 
\end{eqnarray}

For our numerical results showed in Tables I and II we used $\alpha_s = 0.120$, 
$\alpha = \frac{e^2}{4 \pi}= 1/128$, $m_t = 175$ GeV, $m_b = 4.3$ GeV,
$m_c = 1.3$ GeV, $s_w = 0.473$ and 
$M_Z = 91.1867$ GeV.

\end{document}